----------------------------

%




\font\smc=cmcsc10 at 8.77truept
\font\tinyrm=cmr8
\font\verytinyrm=cmr9 at 6truept
\font\tinybf=cmbx8 
\font\tinyit=cmti8
\font\headsmc=cmcsc10 at 6.67truept
\font\headbf=cmbx10 at 7.3truept
\font\headrm=cmr10 at 7.3truept

\thinmuskip=2mu
\medmuskip=3mu plus 1mu minus 3mu
\thickmuskip=4mu plus 4mu

\def\fixcircle{\sevensy\char"0D}
\def\tcopyright{{\ooalign{\hfil\lower.05ex\hbox{C}\hfil\cr\cr\fixcircle}}}

%
%
\def\today{\number\year\space \ifcase\month\or 	January\or February\or 
	March\or April\or May\or June\or July\or August\or September\or
	October\or November\or December\fi\space \number\day}

\newdimen\fullhsize
\fullhsize=7.53truein
\def\fullline{\hbox to\fullhsize}
\def\makeheadline{\vbox to 0pt{\vskip-22.5pt
 \fullline{\vbox to8.5pt{}\the\headline}\vss}\nointerlineskip}
\def\makefootline{\baselineskip=12pt\fullline{\the\footline}}
\def\rightheadline{\rm No.~1,\ \number\year \hfil \shorttitle \hfil L\folio}
\def\leftheadline{\rm L\folio \hfil \shortauthor \hfil Vol.\ 999}
\def\asthead{\vbox{\noindent{\headsmc The Astrosynthetic Journal,
 \headbf 999\headrm :L1--L4, \number\year\ February 31\hfil}\vskip3pt
 \noindent{\verytinyrm\tcopyright 1985, Berkeley
 Astronomy Department. All rights reserved. Printed in U.S.A.\hfil}}}
\headline={\ifnum\pageno=1 \asthead \else{\ifodd\pageno\rightheadline
 \else\leftheadline\fi}\fi}
\footline={\ifnum\pageno=1 {\hfil L1\hfil} \else{\hfil}\fi}

\voffset=-0.05truein
\vsize=9.1truein
\hoffset=-0.55truein
\hsize=7.53truein 
\parindent=.12truein
\parskip=0pt
\baselineskip=10.92truept

\def\beginabstract{\vskip 0pt\centerline{\tinyit Written \today}
 \begingroup\leftskip=.5truein\rightskip=.5truein\parskip=3truept
 \vskip 12pt\centerline{ABSTRACT}}

\def\subjectheadings#1{\vskip 0pt\par\noindent{\it Subject headings:\/} #1}
\def\endabstract{\vskip 6pt\endgroup\begindoublecolumns}
\def\beginref{\enddoublecolumns\medskip\smallskip\centerline
  {\smc references}\begindoublecolumns\begingroup\baselineskip=7.9pt
  \parindent=0in\let\bf=\tinybf\let\it=\tinyit\let\rm=\tinyrm\rm}
\def\endref{\smallskip\endgroup\vfil\enddoublecolumns\parskip=12truept
  \parindent=0in}
\def\section#1{\medskip\penalty -500\centerline{\smc #1}\medskip}
%
%

\def\>{$>$}
\def\<{$<$}

\def\simlt{\lower.5ex\hbox{$\; \buildrel < \over \sim \;$}}
\def\simgt{\lower.5ex\hbox{$\; \buildrel > \over \sim \;$}}
\def\sqr#1#2{{\vcenter{\hrule height.#2pt
      \hbox{\vrule width.#2pt height#1pt \kern#1pt
         \vrule width.#2pt}
      \hrule height.#2pt}}}

%
\def\pp{\par\hangindent=.125truein \hangafter=1}
\def\apjref#1;#2;#3;#4{\pp #1, {\it #2}, {\bf #3}, #4.}
\def\ref#1;#2;#3;#4{\pp #1, {\it #2}, {\bf #3}, #4}
\def\book#1;#2;#3{\pp #1, {\it #2}, #3}
\def\rep#1;#2;#3{\pp #1, #2, #3}

%
%
%
\let\Noline=\relax 
\tolerance=10000
\newdimen\colwidth \newdimen\bigcolheight 
\newdimen\pagewidth \newdimen\pageheight 
%
%
\colwidth=\hsize
  \advance\colwidth by -.35truein
  \divide\colwidth by 2
\bigcolheight=\vsize
  \advance\bigcolheight by \vsize
\newdimen\savevsizea \savevsizea=\vsize \advance\savevsizea by 24pt
\newdimen\savevsize \savevsize=\vsize
\newdimen\savehsize \savehsize=\hsize
\def\makefootline{\baselineskip=24pt\hbox to \savehsize{\the\footline}}
\font\sevenrm=cmr7 at 7truept
\font\fiverm=cmr5 at 5truept
\font\seveni=cmmi7 at 7truept
\font\fivei=cmmi5 at 5truept
\font\sevensy=cmsy7 at 7truept
\font\fivesy=cmsy5 at 5truept
\def\Footstrut{\hbox{\vrule height6.72pt depth1.92pt width0pt}}
\def\sevenpoint{\def\rm{\fam0\sevenrm}
	\textfont0=\sevenrm \scriptfont0=\fiverm
	\textfont1=\seveni \scriptfont1=\fivei
	\textfont2=\sevensy \scriptfont2=\fivesy
	\textfont3=\tenex \scriptfont3=\tenex
	\normalbaselineskip=8.64truept
	\normalbaselines\rm}
\def\footnote#1{\edef\@sf{\spacefactor\the\spacefactor}#1\@sf
  \insert\footins\bgroup\sevenpoint
  \interlinepenalty=\interfootnotelinepenalty
  \let\par=\endgraf
  \splittopskip=\ht\strutbox 
  \splitmaxdepth=\dp\strutbox \floatingpenalty=20000
  \leftskip=0pt \rightskip=\colwidth \advance\rightskip by .35in 
	\spaceskip=0pt \xspaceskip=0pt \parindent=1em
  \indent \bgroup\Footstrut #1\aftergroup\Footstrut\egroup
	\let\next}
\pagewidth=\hsize \pageheight=\vsize
\def\onepageout#1{\shipout\vbox{
    \offinterlineskip
    \makeheadline
    \vbox to\savevsizea{#1
	\boxmaxdepth=\maxdepth}
    \makefootline}
    \advancepageno}
  \output{\onepageout{\unvbox255}}
\newbox\partialpage
\def\begindoublecolumns{\begingroup
  \output={\global\setbox\partialpage=\vbox{\unvbox255}}\eject
  \output={\doublecolumnout} \hsize=\colwidth \vsize=\bigcolheight
  \ifvoid\footins\else\advance\vsize by -\ht\footins\fi
  \advance\vsize by -2\ht\partialpage}
\def\enddoublecolumns{\output={\balancecolumns}\eject
  \global\output={\onepageout{\unvbox255}}
  \global\vsize=\savevsize
  \endgroup \pagegoal=\vsize}
\def\doublecolumnout{\dimen0=\pageheight
  \advance\dimen0 by-\ht\partialpage \splittopskip=\topskip
  \ifvoid\footins\setbox0=\vsplit255 to\dimen0\else
   \dimen1=\dimen0
   \advance\dimen1 by-\ht\footins
   \advance\dimen1 by-12pt
   \setbox0=\vbox to \dimen0{\vss\vsplit255 to\dimen1 
	\vskip\skip\footins \kern-3pt \unvbox\footins}\fi
  \setbox2=\vsplit255 to\dimen0
  \onepageout\pagesofar
  \global\vsize=\bigcolheight
  \unvbox255 \penalty\outputpenalty}
\def\pagesofar{\unvbox\partialpage
   \wd0=\hsize \wd2=\hsize \hbox to\pagewidth{\box0\hfil\box2}}
\def\Makevrule{\gdef\pagesofar{\unvbox\partialpage
  \wd0=\hsize \wd2=\hsize \hbox to\pagewidth{\box0\hfil\vrule\hfil\box2}}}
\def\balancecolumns{\setbox0=\vbox{\unvbox255} \dimen0=\ht0
  \advance\dimen0 by\topskip \advance\dimen0 by-\baselineskip
  \divide\dimen0 by2 \splittopskip=\topskip
  {\vbadness=10000 \loop \global\setbox3=\copy0
    \global\setbox1=\vsplit3 to\dimen0
    \ifdim\ht3>\dimen0 \global\advance\dimen0 by1truept \repeat}
  \setbox0=\vbox to\dimen0{\unvbox1}
  \setbox2=\vbox to\dimen0{\unvbox3}
  \global\output={\balancingerror}
  \pagesofar}
\newhelp\balerrhelp{Please change the page
                        into one that works.}
\def\balancingerror{\errhelp=\balerrhelp
        \errmessage{Page can't be balanced}
        \onepageout{\unvbox255}}
%
\Noline

%
%
%
%
%

\input epsf.sty
\baselineskip=10.92 truept


\def\HST{{\it HST}}
\def\msun{$m_{\odot}$}

\def\simlt{$\lower.7ex\hbox{${\buildrel{<}\over \sim}$}$}
\font\bigbf=cmbx12
\font\smit=cmti8
\font\eightrm=cmr8

\def\shorttitle{LIMIT OF HYDROGEN BURNING}
\def\shortauthor{KING ET AL.}
\null
\vskip 0.30truein
                                
\centerline{THE LUMINOSITY FUNCTION OF THE GLOBULAR CLUSTER NGC 6397}
\centerline{NEAR THE LIMIT OF HYDROGEN BURNING\footnote{$^1$}{Based on
observations with the NASA/ESA {\seveni Hubble Space Telescope}, obtained
at the Space Telescope Science Institute, which is operated by AURA, 
Inc., under NASA contract NAS 5-26555.}}

\centerline{\smc{Ivan R.\ King\footnote{$^2$}{Astronomy 
Department, University of California, Berkeley, CA 94720--3411;
king@glob.berkeley.edu, jay@cusp.berkeley.edu}, Jay Anderson{$^2$},
Adrienne M.\ Cool\footnote{$^3$}{Department of Physics and Astronomy,
San Francisco State University, 1600 Holloway Avenue, San Francisco, CA
94132; cool@sfsu.edu}, and Giampaolo Piotto\footnote{$^4$}{Dipartimento 
di Astronomia, Universit\`a di Padova, Vicolo dell'Os\-ser\-va\-tor\-io 5,
I--35122 Padova, Italy; piotto@astrpd.pd.astro.it}}}


\ \vskip -0.07truein

\beginabstract

Second-epoch \HST\ observations of NGC 6397 have led to the measurement
of proper motions accurate enough to separate the faintest cluster stars
from the field, thus extending the luminosity function of this globular
cluster far enough to approach the limit of hydrogen burning on the main
sequence.  We isolate a sample of 1385 main-sequence stars, from just
below the turnoff down to $I=24.5$ ($M_I\simeq12.5$), which corresponds
to a mass of less than 0.1 \msun\ for the metallicity of this cluster.
Below $I=22$ ($M_I\simeq10$), the main-sequence luminosity function
drops rapidly, in a manner similar to that predicted by theoretical
models of low-mass stars near the hydrogen-burning limit.

\subjectheadings{Globular Clusters: Individual (NGC 6397) --- Stars:
Hertzsprung--Russell Diagram --- Stars: Interiors --- Stars: Low-Mass,
Brown Dwarfs}
\endabstract


\section{1. Introduction}

On the lower main sequence, there is a mass limit below which the
center of a contracting protostar fails to reach the temperature
needed for the hydrogen-burning nuclear reactions that characterize
the life of a main-sequence star.  At this limit of hydrogen burning,
the stars of an otherwise smooth distribution of stellar masses divide
between normal main-sequence stars, above, and brown dwarfs, below.
The former shine continuously for much longer than a Hubble time,
while the latter fade, so that after the $>$10 Gyr age of a globular
cluster has passed, brown dwarfs are well separated in luminosity from
hydrogen-burning stars (e.g., Burrows et al.\ 1993).  Among the
hydrogen-burning stars in the near neighborhood of the limit, theory
predicts a steep decline in the luminosity function (LF), even before
the more abrupt plunge of the LF when the limit is reached (e.g.,
D'Antona 1995, and other sources to be discussed below).

This limit should manifest itself particularly well in globular
clusters, for three reasons: (1) The stars of a cluster have a common
age and composition, and can be observed as a compact group.  (2) At the
low metal abundances of globular clusters, the hydrogen-burning limit
occurs at a higher mass than for stars of solar metal abundance.  (3)
Stars of low metallicity are more luminous, at a given mass, than are
those of high metallicity.  It was for these reasons that Vittorio
Castellani and Vittoria Caloi proposed to the European Space Agency in
1984 that the Hubble Space Telescope be used to look for the lower limit
of hydrogen burning on the lower main sequences of the nearest globular
clusters.

The cluster of smallest distance modulus is NGC 6397, for which we adopt
$(m-M)_I \simeq 12.05$.  Two WFPC2 observing programs have been directed
at the faint stars of this cluster (Paresce, De Marchi, \& Romaniello
1995; Cool, Piotto, \& King 1996 = CPK).  Both studies showed a drop in
the luminosity function below $I=21$, but with weak statistics.  As CPK
emphasized, what was lacking was not a faint limiting magnitude, but
rather the ability to distinguish the faintest cluster stars from the
much more numerous field stars.

On encountering this frustration, however, we realized that the proper
motion of NGC 6397 ($\simeq15$ mas/yr [Cudworth \& Hanson 1993]) is
large enough that a repeat of our 1994 exposures should allow a
proper-motion separation of individual cluster members from field stars.
We therefore re-imaged the cluster in 1997.  We expected a mean cluster
motion of over 0.4 pixel, with the internal velocity dispersion of the
cluster contributing a little over 0.01 pixel scatter.

\section{2. Observations}

For various observational reasons, the 1994 exposures had covered an
irregular pattern (CPK, Fig.\ 1).  For the second-epoch we chose two
pointings that maximize the number of stars covered by at least three
orbits at each epoch.  They include about 88\% of the 2770 stars
measured by CPK.  The second-epoch exposures were made with the F814W
filter on June 4, 1997.

The first-epoch exposures had been dithered by circumstance; those of
the second epoch were deliberately placed at three different
fractional-pixel positions within each of the two major pointings.  Each
dither position was held for one full orbit, which was divided into
three exposures of 700--800 s each, and one short exposure of 60 s.

Individual images were cleaned of cosmic rays by intercomparison with
one another, and the long exposures of each orbit were then averaged.
The result was six single-orbit images, consisting of three dithers at
each of two pointings.

The basic aim of the astrometry was to establish which of the faint
stars shared the motion of the cluster and which did not.  For each star
measured in the first epoch, we used the first- and second-epoch
positions of the nearest ten or so bright cluster members (i.e, stars
with $I<22$ and $V-I$ within 0.05 mag of the main-sequence ridge line),
to determine the presumed location of the target star in each of the
second-epoch images.

We then searched for a star at this position.  If the star had the same
proper motion as the cluster stars, we would expect to find it in the
predicted location.  Any difference of motion from that of the cluster
should manifest itself by an offset from this location.  We measured
positions for the brighter stars ($I \simlt 24$) by least-squares
PSF-fitting.  For faint stars we determined a center from a {\bigbf
+}-shaped set of 5 pixels centered on the star's brightest pixel, using
a PSF-based algorithm that will be described in a later paper.  The
level of accuracy of our astrometry will be indicated by the scatter in
the measured motions of the cluster stars, all of whose true motions
should be nearly zero in the system described.

\medskip

\epsfxsize=88truemm\epsfbox{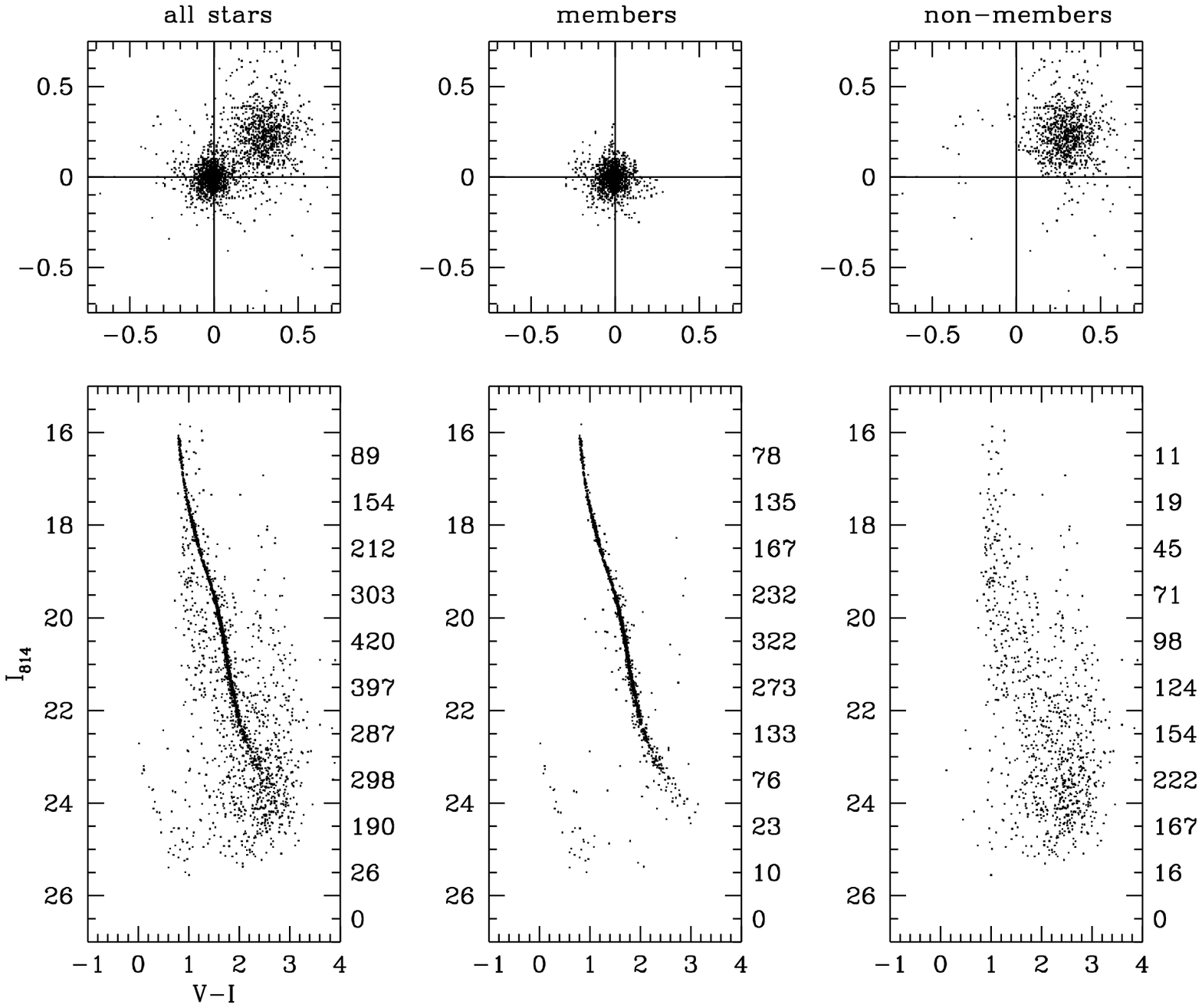}
\ \vskip -18truemm

{\eightrm {\baselineskip=9pt Figure 1. Proper-motion distributions,
above, and color--magnitude diagrams, below.  The scale of the proper
motions is displacement in WFC pixels over the 32-month time baseline;
a full WFC pixel of displacement would correspond to 37.5 mas/yr.  Since
all reference stars were cluster members, the zero point of motion is
the mean motion of cluster stars.  Left: the entire sample; center:
stars within the proper-motion region described in the text; right:
stars outside this region.  Numbers at right are stars per
unit-magnitude bin.\bigskip}}

\section{3. Results and Discussion}

In Figure 1 we show the distribution of proper motions and the
color--magnitude diagram (CMD) for the entire sample, as well as for two
subsets of stars, divided according to whether the proper motion of a
star falls inside or outside of a boundary in the proper-motion diagram
that approximately isolates the cluster members.  For this boundary we
chose a radius of 0.15 pixel in the quadrant that contains most of the
field stars and 0.3 pixel in the other three quadrants.  These radii
were chosen by trial and error so as to achieve the best easy separation
between cluster members and field stars.

The success of the separation is immediately evident.  It is not
perfect, of course, because (1) the error distribution causes some
cluster members to fall outside the circle, and (2) the proper motions
of the field stars fall all over the graph, so that inevitably some
field stars have motions that are close to that of the cluster.

In the cluster-member CMD in Figure 1 the number of cluster stars is
dropping off sharply below $I=23.5$ (see also Fig.\ 2), and this is not
purely incompleteness, since the non-member set is full of stars at
least to $I=24.5$.  In fact, the completeness figures are already known,
since the stars studied here are those of the first-epoch study (CPK),
where thorough completeness tests were carried out.  At $I=23.5$ our
completeness is 90\%, and it is still 78\% at $I=24.5$ ({\it cf.} Table
1).  As nearly every epoch-1 star within the field of view of the
second-epoch observations was recovered (see below), these completeness
figures still apply.

\epsfxsize=88truemm\epsfbox{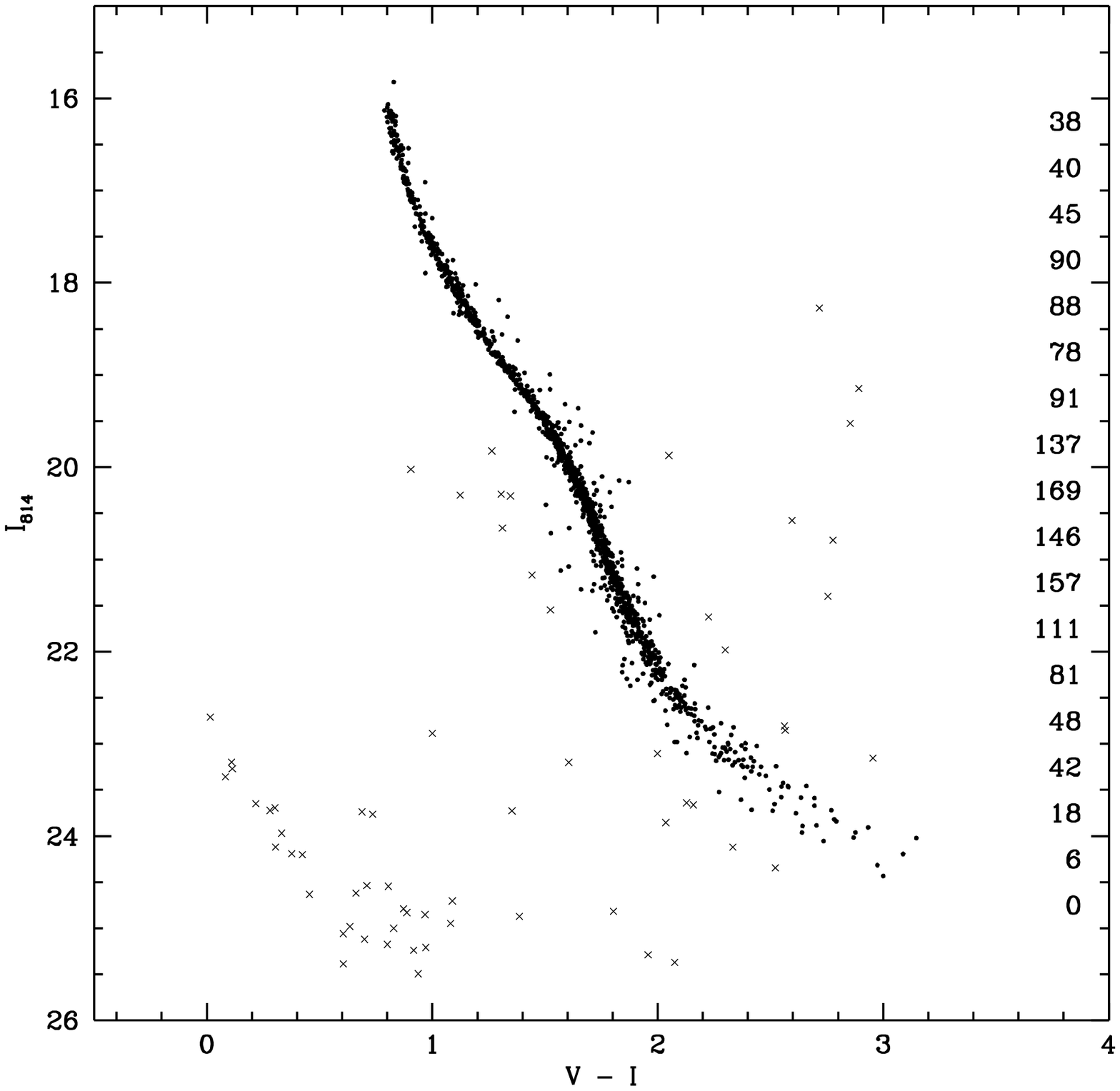}

{\eightrm {\baselineskip=9pt Figure 2. Blow-up of bottom center panel of
Fig.\ 1, with counts per half-magnitude interval tallied at the right.
Dots:\ stars within 0.2 mag of the color of the main-sequence ridge line
(only these are counted in the luminosity function); crosses:\ stars
outside this color range.\bigskip}}

Figure 2 is a blowup of the CMD of the cluster sample in Fig.\ 1; stars
whose colors fall within 0.2 mag of the main-sequence ridge line are
plotted as filled circles and are considered to be cluster members.  For
the brighter stars this color criterion is overly generous and errs in
the direction of allowing some field stars into the sample.  (It also
allows for the possibility of a population of cluster binaries to the
right of the main sequence---a question to which we shall return in a
later paper.)  For the fainter stars, however, where the photometric
errors increase, it insures that cluster stars are not lost outside a
color range that is too narrow.  Because of the preponderance of field
stars at the faint end, even though an occasional cluster star may be
missed there, the net error goes in the direction of making the drop-off
of the number of cluster stars {\it less} steep.

Another concern is the possibility that faint cluster members are lost
astrometrically, as an increasing measurement error pushes them beyond
the proper-motion separation criterion.  We examined the overall
distribution of proper motions as a function of magnitude.  Figure 3
shows the inner part of this distribution in the faintest four 1-mag
intervals.  The spread of apparent proper motions of cluster stars does
increase somewhat with faintness, but very few of them should be lost by
the separation criterion that was used.  In the $I=24$ to 25 interval,
we examined each individual star that lies between the solid arc and the
dashed arc (at radius 0.25) in the first quadrant; two of the eight
actually fall just within our color limits, but the color distribution
of the eight is spread in a way that is consistent with these two being
just the red tail of a field-star distribution.

\epsfxsize=88truemm\epsfbox{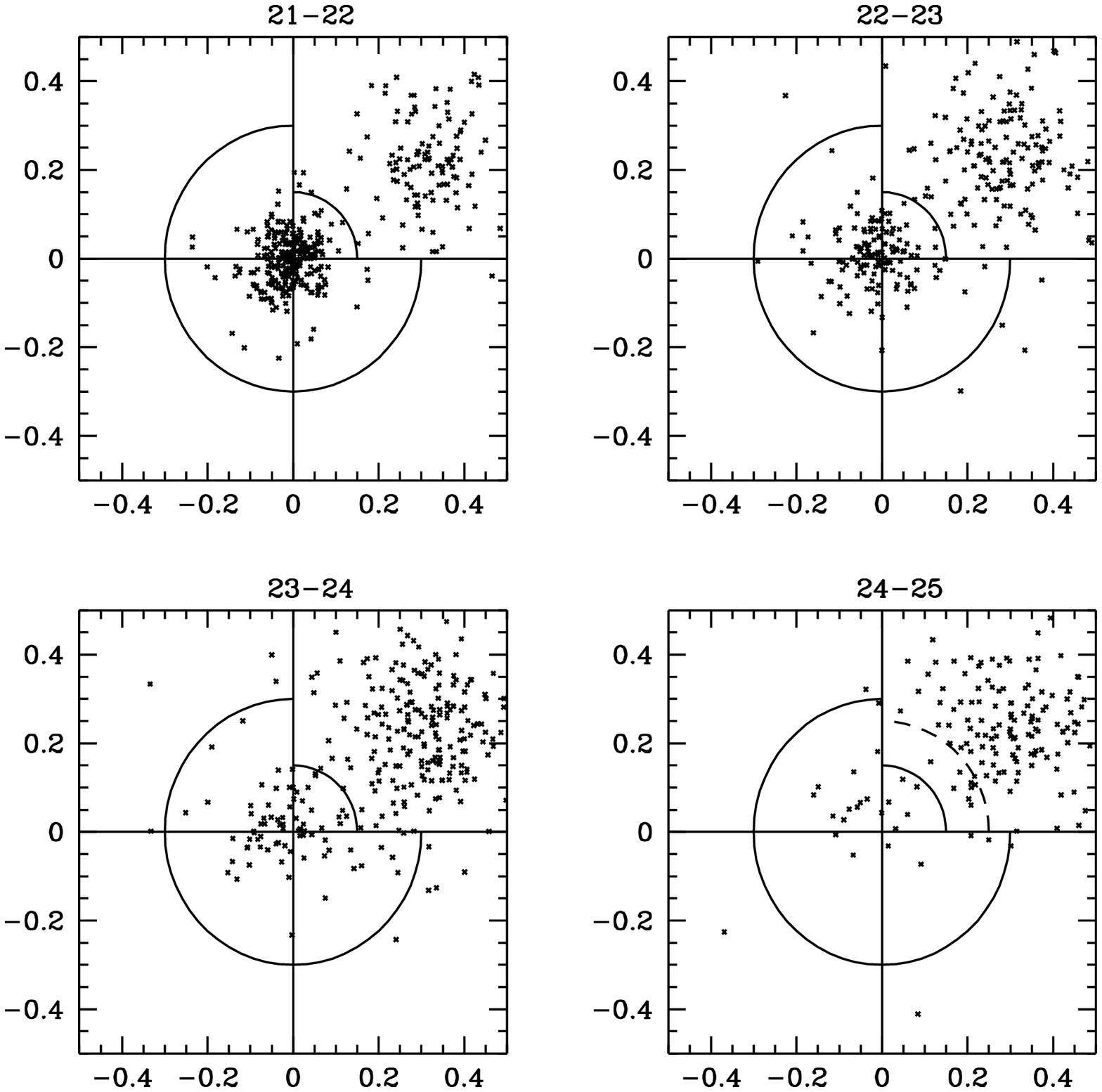}

{\eightrm {\baselineskip=9pt Figure 3. Proper motions in each of our
faintest four unit intervals of {\seveni I} magnitude.  The arcs show our
separation criterion between cluster and field; the dashed arc is
explained in the text.\bigskip}}

{\parfillskip=0pt
In order to convert the numbers at the right edge of Fig.\ 2 to a
luminosity function, we need careful completeness corrections. 
Our present list is the stars of the first-epoch study, whose
completeness was studied by CPK.  In Table 1 we give the full results of
the completeness tests made for that study.  In principle, an additional
correction might \vskip 0pt}

\ \vskip 2truemm

{\eightrm {\baselineskip=9pt
\centerline{Table 1. Completeness.}
\ \vskip -7truemm

$$\vbox{\halign {\hfil # \hfil & \hfil # \hfil & \qquad \qquad # \hfil 
& \hfil # \hfil & \hfil # \hfil \cr
\noalign{\bigskip\hrule\smallskip\hrule\medskip}
 {\smit I} &  {\eightrm Fraction} &&  {\smit I} & {\eightrm Fraction} \cr
\noalign{\medskip\hrule\smallskip}
 16.50 &  1.00 &&  23.50 & 0.90 \cr
 17.50 &  1.00 &&  24.00 & 0.87 \cr
 18.50 &  1.00 &&  24.25 & 0.85 \cr
 19.50 &  0.98 &&  24.50 & 0.78 \cr
 20.50 &  0.97 &&  24.70 & 0.68 \cr
 21.50 &  0.96 &&  24.90 & 0.45 \cr
 22.50 &  0.93 &&  25.00 & 0.35 \cr
}}$$
\bigskip}}

\noindent be needed for stars missed in the second-epoch
astro\-metric study, where we included only stars that were identified and
measured in all three orbits of at least one of the two pointings.  But
this loss was almost purely geometric (the coverage was 88\% of the
previous field); in the area covered by the second-epoch images, only 27
stars out of 2438 were missed, and these were distributed fairly evenly
over magnitude.

Our new LF is plotted in the upper part of Figure 4.  It agrees well
with with the earlier CPK points, except for the last of the latter,
which was affected by the very field contamination that we are removing
here.  We note that we showed in a previous paper (King, Sosin, \& Cool
1995) that the field we have studied is at a distance from the cluster
center where, fortuitously, the local LF should be very similar to the
cluster's global LF.

\epsfxsize=88truemm\epsfbox{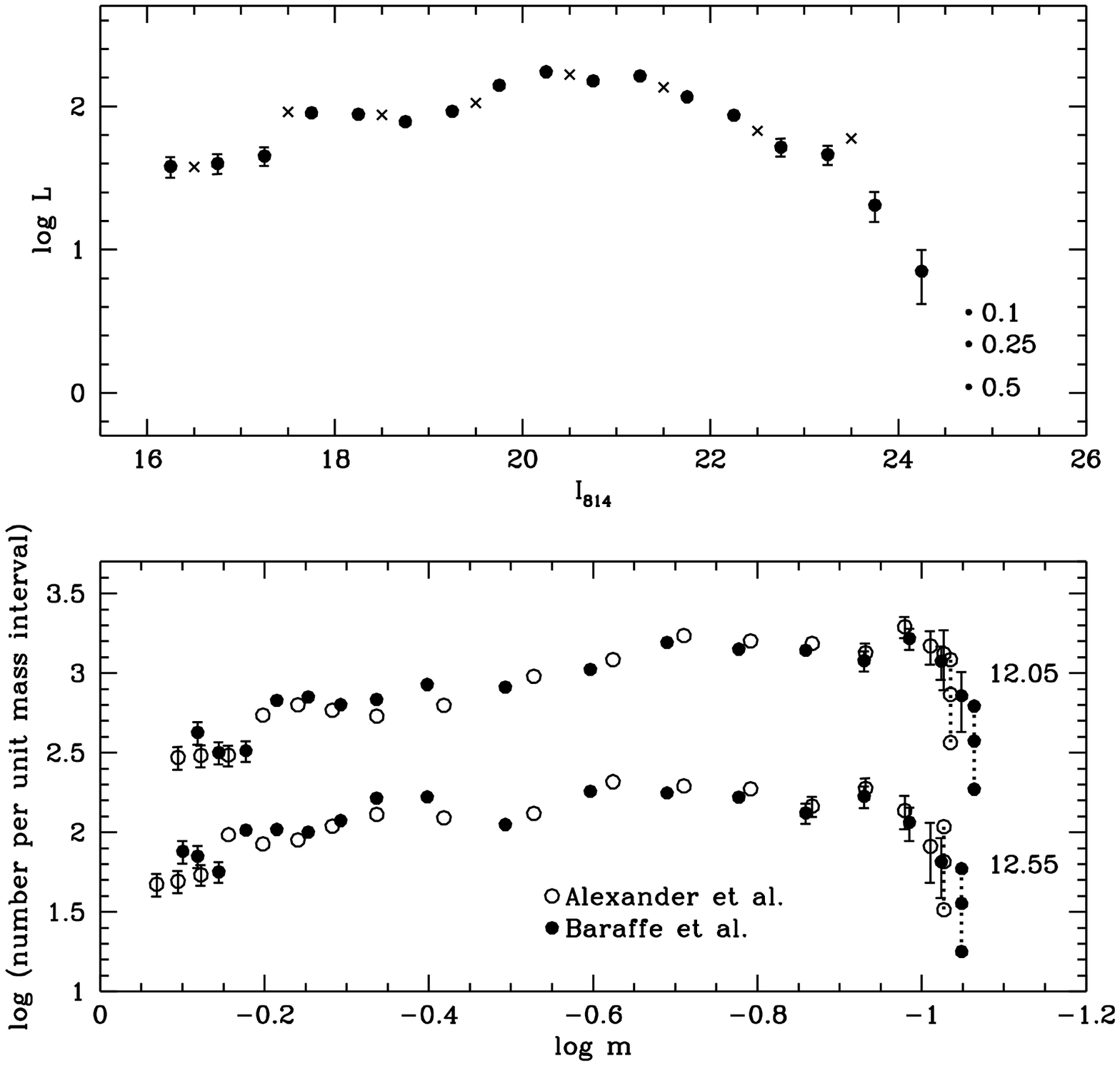}

{\eightrm {\baselineskip=9pt Figure 4. (top) Our new luminosity function
of NGC 6397, with Poisson error bars (plotted only when they are larger
than the sizes of the symbols).  The vertical array of three small dots
is explained in the text.  The crosses are the LF given by CPK,
converted to the present field size.  (bottom) Mass functions, as
derived from each of the two MLRs indicated.  For clarity the sets of 3
points representing the empty bin have been connected with lines.  The
error bars arise from those of the log LF points.  The MFs are shown for
two different assumed distance moduli, as labeled.\bigskip}}

A problem is posed, in plotting the LF, by the fact that our faintest
half-magnitude bin has no observed cluster stars; the statistical
interpretation of a zero count is always a problem.  What we have done
is to ask, for various putative values of the LF at $I=24.75$, what the
probability would be of our observing a count of zero in this bin.  The
points plotted are such that if the true value of the LF were at that
point, the probability of observing zero stars in that bin would be,
respectively, 0.1, 0.25, and 0.5, for the three points plotted.  The
results suggest that it is quite unlikely that this bin fails to
accentuate the steep plunge of the LF.

We should caution against over-interpreting this zero-count bin,
however.  As Table 1 shows, over the magnitude range of this bin
($I=24.5$ to 25.0) the completeness of the sample drops from 0.78 to
0.35.  We have consequently been re-examining our images, to see if we
can possibly find additional main-sequence cluster members in this
range.  This study will not be complete for some time, but it seems not
unlikely that our field will yield up a star or two that is below
$I=24.5$ and is on the cluster main sequence.  Even so, the LF would be
dropping {\it very} rapidly at this magnitude.

We now turn to theoretical studies of stellar structure, in order to
see how the drop-off in our new LF compares to what is expected just
above the limit of hydrogen burning, from current models of these
low-metallicity stars ([Fe/H] $\simeq-1.9$ [Djorgovski 1993]).  We do
this by using theoretical mass--luminosity relations (MLRs) to derive
the mass function (MF) that is implied by our LF.  In principle, if
the MLR is correct, the resulting MF should behave smoothly near the
H-burning limit, because the process of star formation should be
unaware of the later restrictions imposed by the physics of energy
generation.  Though fundamentally qualitative, this smoothness
criterion is a potentially sensitive test.  Because of the way that the
LF maps into the MF, there will be an abrupt feature in the MF unless
the steep drop in the LF is matched by a comparably steep rise in the
MLR slope, by which the LF is multiplied in order to convert it into an
MF.

We have derived MFs using two values of the $I$ distance modulus, and
two different MLRs.  The latter are derived from stellar models which
have been shown by their authors (Alexander et al.\ 1997, Baraffe et
al.\ 1997) to give a satisfactory fit to the CPK main-sequence ridge
line in NGC 6397.  In examining the MFs, however, it should be kept in
mind that these two sets of models are characterized not only by
different underlying physics, but also by different assumed chemical
compositions.  As for the distance modulus, because its value controls
the alignment of the theoretical slopes with the observed points, it can
have a significant impact on the resulting MF (D'Antona 1998).  The
distance moduli of 12.05 and 12.55 that we have chosen bracket the range
in which the true value is likely to lie.

The results are shown in the lower half of Fig.\ 4.  The broad features
of the MFs are similar in all cases, rising gradually and then leveling
off at low masses.  There are indications of features in the MFs near
$\log m = -.015$--0.2 and $-0.45$.  The former region, however, is close
to the main-sequence turnoff, where the MLR is sensitive to age and
evolution.  Both features will be discussed elsewhere, along with the
extensive LF material that is available on other clusters (as, for
example, in Piotto, Cool, \& King [1997], where we give our previous LF
values along with those for three other clusters).

Here our focus is on the low-mass region.  It is noteworthy that the
striking drop at the faint end of the luminosity function is essentially
absent in the MF.  The steepness of the theoretical mass--luminosity
relation near the H-burning limit has largely compensated for it,
raising the MF up to be nearly flat below $\log m \sim -0.7$.  In the
last two occupied bins we see some indication of a turn-down, with the
empty bin accentuating this effect.  This takes place within a mass
range of only a little over 0.01 \msun.  But because the amount of
turn-down hardly exceeds the size of the error bars, and because of the
even larger uncertainty associated with the empty bin, the present study
cannot determine with any confidence whether or not the MF begins to
drop below $\log m \sim -1.0$.  Any sudden turn-down of the MF within a
small mass range would be surprising; it would be of great interest if
it were a real feature of the MF, or it could equally well indicate that
the MLRs do not have a steep enough slope in this region.  Thus improved
data for the now-empty last bin could be quite significant.  Also needed
are study of a larger sample within NGC 6397, and examination of other
clusters of small distance modulus.

\section{4. Summary}

The main-sequence luminosity function that we present for NGC 6397
reaches into the mass range immediately above the limit of hydrogen
burning.  Our results demonstrate that \HST\ proper motions can be
very effective in distinguishing cluster stars from the field.  The
newly derived LF is no longer subject to the uncertainties associated
with field stars that limited previous studies of this cluster, and
extends to faint enough magnitudes to include stars with masses below
0.1 \msun.  We use mass--luminosity relations from theoretical studies
to convert our LF into an MF, and find that the resulting MF is
reasonably smooth down to 0.1 \msun.  Statistical uncertainties limit
what can be gleaned about the MF at even lower masses,

In future months we will refine all of the measurements and calculations
that have gone into this brief Letter, in order to present as accurate a
delineation of the lower main sequence of NGC 6397 as we can.  We
believe, however, that even the present preliminary results are of
interest, showing for the first time the neighborhood of the lower limit
of hydrogen burning on the main sequence of a globular cluster.

\bigskip

We are grateful to the Teramo group (Alexander at al.\ 1997) and the
Lyon group (Baraffe et al.\ 1997) for communication of data prior to
publication.  We also gratefully acknowledge helpful discussions of the
stellar models with Francesca D'Antona, Lars Bildsten, Santi Cassisi,
and Gilles Chabrier.  This work was supported by NASA grant GO--6797
from STScI (IRK, JA, and AMC), as well as by the Agenzia Spaziale
Italiana and the Ministero dell'Universit\`a e della Ricerca Scientifica
e Tecnologica (GP).

\beginref

Alexander, D.\ R., Brocato, E., Cassisi, S., Castellani, V.,
Ciacio, F., \& Degl'Innocenti, S. 1997, A\&A, 317, 90
 
Baraffe, I., Chabrier, G., Allard, F., \& Hauschildt, P.\
H. 1997, A\&A, in press (Nov.\ 1997)

Burrows, A., Hubbard, W.\ B., Saumon, D., \& Lunine, J.\
I. 1993, ApJ, 406, 158

Cool, A.~M., Piotto, G., \& King, I.~R. 1996, ApJ, 468, 655.

Cudworth, K.\ M., \& Hanson, R.\ B. 1993, AJ, 105, 168

D'Antona, F. 1995, in The Bottom of the Main Sequence --- and
Beyond, ed.\ C.\ G.\ Tinney (Berlin:\ Springer), p.\ 13
 
D'Antona, F. 1998, in The Stellar Initial Mass Function, ed.\ 
G.\ Gilmore, ASP Conference Series, in press

Djorgovski, S.\ G. 1993, in Structure and Dynamics of
Globular Clusters, eds.\ S.\ G.\ Djorgovski \& G.\ Meylan (San
Francisco:\ ASP), p.\ 373

Holtzman, J., et al.\ (18 authors). 1995, PASP, 107, 156

King, I.\ R., Sosin, C., \& Cool, A.\ M. 1995, ApJ, 452, L33

Paresce, F., De Marchi, G., \& Romaniello, M. 1995, ApJ, 440, 216 
 
Piotto, G., Cool, A.\ M., \& King, I.\ R. 1997, AJ, 113, 1345.

\endref

\end